\documentclass[11pt]{article}
\usepackage{fancyhdr}
\usepackage{isomath}
\usepackage{amsmath}
\usepackage{amsbsy}
\usepackage{amssymb}
\usepackage{amscd}
\usepackage{amsfonts}
\usepackage{graphicx,color}
\usepackage{verbatim}
\usepackage{euscript}
\usepackage{alltt}
\usepackage{stmaryrd}
\usepackage{subfigure}
\usepackage{xurl}
\usepackage{hyperref}
\usepackage{comment}

\newenvironment{rcases}
  {\left.\begin{aligned}}
  {\end{aligned}\right\rbrace}

\newcommand{\veps}{\varepsilon}
\newcommand{\p}{\partial}
\newcommand{\dee}{\mathcal{D}}
\newcommand{\scl}{\mathcal{L}}

\usepackage[margin=1in]{geometry}

\date{}

\begin{document}
\title{Action principles for dissipative, non-holonomic Newtonian mechanics}
\author{Amit Acharya\thanks{Department of Civil \& Environmental Engineering, and Center for Nonlinear Analysis, Carnegie Mellon University, Pittsburgh, PA 15213, email: acharyaamit@cmu.edu. Corresponding author.} $\qquad$ Ambar N. Sengupta\thanks{Department of Mathematics, University of Connecticut, Storrs, CT 06269, email: ambarnsg@gmail.com} }

\maketitle

\begin{abstract}
\noindent A methodology for deriving dual variational principles for the classical Newtonian mechanics of mass points in the presence of applied forces, interaction forces, and constraints, all with a general dependence on particle velocities and positions, is presented. Methods for incorporating constraints are critically assessed. General theory, as well as explicitly worked out variational principles for a dissipative system (due to Lorenz) and a system with anholonomic constraints (due to Pars) are demonstrated. Conditions under which a (family of) dual Hamiltonian flow(s), as well as a constant(s) of motion, may be associated with a conservative or dissipative, and possibly constrained, primal system naturally emerge in this work.
\end{abstract}

\section{Introduction}\label{sec:intro}
In this paper, we develop a variational/action principle for Newton's equations of motion for a system of particles possibly subjected to kinematic constraints which are nonlinear in the particle velocities, and forces that may not arise from a potential function of the particle positions. Along the way, we also critically consider the various prevalent formulations in the literature of constraint forces. Although a classical subject, the subtleties involved, even in the setting of anholonomic, linear-in-velocities constraints, and the nature of unresolved issues can be appreciated from the works \cite{cronstrom2009nonholonomic,lewis1995variational}. The primary applications that we envisage of such variational principles is to facilitate path integral based statistical formulations of dissipative particle systems and for computing periodic orbits of general ODE systems displaying bounded long-time behavior.

A scheme for generating action principles for dissipative initial value problems (ordinary differential equations (ODE)) of classical physics has been considered by \cite{galley2013classical}. A special family of Hamiltonians for thermostatted molecular dynamic systems with conservative applied forces is proposed and reviewed in \cite{dettmann1996hamiltonian}, \cite[Sec.~IV]{morriss1998thermostats}. These works are different in concept and scope from ours, which is also applicable to partial differential equations (PDE) \cite{action_2,action_3} and related to the line of thought advanced in \cite{brenier2018initial,brenier_book}. We also mention at the outset that the proposed methodology is different from a `Least-Squares' approach, and in no way hinges on the proposed functional attaining a particular value for its validity; the essential idea of our method can be understood in the finite dimensional setting, as explained in \cite[Sec.~2]{action_3}. For a review of the classical mechanics-related aspects of our work, see \cite{delphenichgauss}.

An outline of the paper is as follows: in Sec.~\ref{sec:intro_constr} we critically discuss the assumptions behind the formulation of constrained particle mechanics. Sec.~\ref{sec:dual_vp} details the main achievements of the work before presenting the fundamental formalism for producing dual variational principles for conservative and/or dissipative classical mechanics with general nonlinear constraints on positions and velocities. This includes an explicitly worked out example on the dissipative Lorenz system of ODE as well as remarks on potential use of our methods for computing periodic orbits of general ODE systems. Sec.~\ref{sec:gen_Pars} continues with further demonstration of our scheme in the setting of a key system with anholonomic constraints due to Pars \cite{pars1954variational}, and its generalization. Sec.~\ref{sec:resol} contains a proposal, free of ad-hoc assumptions, for resolving the non-uniqueness in the form of the constraint forces in particle systems with nonlinear constraints. Finally, Sec.~\ref{sec:concl} contains some concluding remarks.

Holonomic systems were identified and studied by Hertz \cite{Hertz1910}. Non-holonomic systems, though realistic and common in practice, have been discussed in the literature over the years (\cite{flannery2005enigma, cronstrom2009nonholonomic,BMZ2004} to mention a few), but there seems to be no standardized treatment that goes beyond constraints linear in velocities. In the latter case, there is the coincidental advantage that the velocities of the solution path satisfy the fixed-time variation constraints. Griffiths \cite{Griff1983} has developed a  framework   that realizes the variational equations for a given Lagrangian, subject to constraints, as describing integral curves of exterior differential systems.   Our task is different, in that we start with the equations of motion and develop a (corresponding family of) Lagrangian(s) (\ref{E:Shdual}) purely on the dual space. The Euler-Lagrange equations of the corresponding action functionals are the given equations of motion, interpreted through a mapping or a change of variables, of the dual variables and their first order derivatives, defining the primal variables appearing in those equations of motion. An additional feature of our method is the role played by a function $H$ that effectively parametrizes classes of solutions to the primal problem.

\section{On the definition of forces maintaining constraints}\label{sec:intro_constr}
It will suffice for us to consider all particle positions, velocities, and forces to be expressed in components w.r.t a fixed rectangular Cartesian coordinate system. A superposed dot will represent a time derivative and two dots, the second time derivative. A twice repeated index in a monomial will indicate summation over the range of the index, if unaccompanied by an explicit summation symbol.

Consider a mechanical system of $N$ particles without any kinematic constraints, each subjected to a net force; with $x_A=(x_{1A},\ldots, x_{dA})\in {\mathbb R}^d $ denoting the configuration coordinates of the $A$-th particle, and $f_A=(f_{1A},\ldots, f_{dA})$ denoting the force divided by the mass of the $A$-th particle, the equation of motion for the system is:
\[
\ddot{x} = f; \qquad (\ddot{x}_{iA} = f_{iA}), \quad i = 1, \dots, d; A = 1, \dots, N,
\]
where $x=(x_A)_{A=1,\ldots, N}$ and $f=(f_A)_{A=1,\ldots, N}$, and, as noted, we have absorbed the particle masses into the  `forces' (which now have physical dimensions of Force/Mass = acceleration) $f_A$. For simplicity, it suffices for this preliminary discussion to even consider the forces $f$ as a given functions of time. Suppose now the system is subjected to a single holonomic kinematic constraint given by
\[
g(x) = 0,
\]
where $g:{\mathbb R}^{dN}\to {\mathbb R}$ is a smooth function.
Then a physically reasonable proposition, and one that is universally accepted, is that the constrained dynamics is defined by the system
\begin{subequations}\label{eq:constrained_disc}
\begin{align}
 \ddot{x} & = f - f^{(c)}; \qquad \left(\ddot{x}_{iA} = f_{iA} - f^{(c)}_{iA} \right), \label{eq:constrained_disc_1}\\
 0 & = g(x), \label{eq:constrained_disc_2}
 \end{align}
\end{subequations}
where $f^{(c)}$ is the \textit{constraint force} that arises to maintain the kinematic constraint(s) \eqref{eq:constrained_disc_2}. As such, the task of the theory of Newtonian particle mechanics is to determine a trajectory
\[
t \mapsto x(t)\in{\mathbb R}^{dN},
\]
with $t$ running over an interval in ${\mathbb R}$ containing the initial time $0$,
that satisfies \eqref{eq:constrained_disc}, subject to prescribed initial conditions
\[
x(0) = x^0; \qquad \dot{x}(0) = v^0.
\]
Uniqueness of solutions $t \mapsto x(t)$ satisfying \eqref{eq:constrained_disc}, at least for short times, is desirable. In the process, it is also desirable to have an idea of the constraint forcing 
\[
t \mapsto f^{(c)}(t)
\]
required in the problem.

It is clear that the above system is formally underdetermined as it has, in general, more variables than equations. Differentiating \eqref{eq:constrained_disc_2} w.r.t time along a trajectory yields
\begin{equation}\label{eq:constr_rate}
    \nabla g (x(t)) \cdot \dot{x}(t) = 0,
\end{equation}
and if one now invokes the additional requirement that the constraint forces expend no power along actual trajectories, i.e., for functions $x(\cdot)$ that satisfy the equations of motion and constraints
\begin{equation}\label{eq:constr_no_work}
   f^{(c)} \cdot \dot{x} = 0, 
\end{equation}
then it is clear that for $N \geq 1, d > 2$, while
\begin{equation}\label{eq:constr_parll}
    f^{(c)}(t) : = \lambda(t) \, \nabla g(x(t))
\end{equation}
for $t \mapsto \lambda(t) \in \mathbb{R}$ is a \textit{sufficient condition} for the satisfaction of \eqref{eq:constr_no_work}, it is certainly not necessary.

If it is demanded that the requirement that `constraints do no work' \eqref{eq:constr_no_work} hold for \emph{all} possible trajectories consistent with only the constraint and not necessarily \eqref{eq:constrained_disc_1} (this is also a consequence of a constrained Hamiltonian description of the dynamics when the applied forces, $f$, arise from a potential), then \eqref{eq:constr_parll} becomes a necessary condition, but it is to be realized that this is, in essence, an extraneous physical assumption as it seems physically natural to demand/know that a constraint force arises to maintain the constraint only along an actual trajectory that also satisfies Newton's second law \eqref{eq:constrained_disc_1}.

 Of course, the above assumption \eqref{eq:constr_parll}, while physically dubious in the above sense, has the major advantage that it immediately makes the dynamical description at least formally determined, that is having the right number of equations and unknowns.

Viewed in the above manner, the argument remains essentially unchanged in the case of anholonomic, homogeneous constraints linear in the velocities characterized by
\begin{equation}\label{eq:anhol_dalembert}
    a(x) \cdot \dot{x} = 0,
\end{equation}
where the ${\mathbb R}^{dN}$-valued function $a$, defined on the configuration space ${\mathbb R}^{dN}$,   takes the place of $\nabla g$, and it is d'Alembert's \textit{principle}, an assumption, that the constraint force is required to do no work on \textit{all} trajectories consistent with the constraint \eqref{eq:anhol_dalembert} (not necessarily satisfying \eqref{eq:constrained_disc_1}). The above arguments also hold, in essence, in the presence of more than one constraint, involving then the same number of Lagrange multiplier fields $\lambda$ (the treatments are standard and can be found in any text book on Analytical Dynamics (cf.~\cite{rosenberg1991analytical, goldstein1957classical}). 

Within this context, it is interesting to assess the controversy regarding the treatment of linear-in-velocity, homogeneous anholonomic constraints by d'Alembert's principle and by the standard Lagrange multiplier rule of the Calculus of Variations, as described in \cite{cronstrom2009nonholonomic, flannery2005enigma}. It suffices to discuss the case of a single anholonomic constraint given by
\begin{equation}\label{eq:anhol_adxdalembert}
    a(x) \cdot \dot{x} = 0.
\end{equation}
The Calculus of Variations treatment gives the constraint force (see, e.g., \cite{cronstrom2009nonholonomic}),  written in terms of generalized coordinates as (and, which, in this specific context can be written in terms of $(A,i) \mapsto \alpha(i,A) = (A-1)d+ i$, $x_{ia} = q_{\alpha(i,A)}$, e.g.)
\begin{equation}\label{eq:cov}
    f^{(c)}_\alpha = - \dot{\mu} \,a_\alpha(q) - \mu (\partial_{ \beta} \,a_\alpha - \partial_{\alpha } \,a_\beta) \dot{q}_\beta,
\end{equation}
whereas the constraint force from the d'Alembert treatment could be interpreted simply as
\begin{equation}\label{eq:covdA}
 f^{(c)}_\alpha = - \dot{\mu} \,a_\alpha(q).   
\end{equation}

Cronstr\"om and Raita \cite{cronstrom2009nonholonomic} prove that the solutions of the dynamical equations using these two descriptions are not identical for identical, prescribed initial conditions. While it is certainly possible to argue the (de)merits of any particular description, we simply note that the constraint force description \eqref{eq:cov} does satisfy all requirements of the d'Alembert specification, namely, that all trajectories consistent with the constraint $a_\alpha \dot{q}_\alpha = 0$ expend no power when acted on by the constraint force.

This further underlines the indeterminacy inherent in the constrained initial value problem (ivp) of analytical dynamics, even under the best of situations when the corresponding unconstrained problem may have unique solutions to the ivp.

In the same spirit, of obtaining equations of motion for constrained systems that have equal number of unknowns and equations by making special assumptions, is the principle of Least Constraint of Gauss \cite{Gauss1829}, as clearly explained in \cite[Sec.  II]{evans1983nonequilibrium} (also see, \cite{delphenichgauss}). Consider a constraint with a completely general, nonlinear in the velocities, dependence of the form
\begin{equation}\label{eq:gen_constr}
    n(x, \dot{x}, t) \cdot \ddot{x} + w(x, \dot{x}, t) = 0,
\end{equation}
where $n$ and $w$ are smooth real-valued functions on ${\mathbb R}^{dN}\times {\mathbb R}^{dN}\times {\mathbb R}$;
this encompasses, holonomic and anholonomic constraints considered earlier. The constraint applies to any trajectory satisfying the equation of motion \eqref{eq:constrained_disc_1}; hence,
\[
n \cdot \left(f - f^{(c)} \right) + w = 0 \Longrightarrow n \cdot f^{(c)} =  (n \cdot f + w).
\]
\emph{Assuming} the constraint force to be of the form
\[
f^{(c)} :=  \lambda \, n
\]
for $\lambda$ a scalar valued function, one obtains the representation
\begin{equation}\label{eq:gauss_lc}
    f^{(c)} =  \frac{(n \cdot f + w)}{n \cdot n}  \, a.
\end{equation}
It can be checked that a trajectory satisfying \eqref{eq:constrained_disc_1} with the prescription \eqref{eq:gauss_lc} also satisfies the constraint \eqref{eq:gen_constr}. Thus, Gauss's principle of Least Constraint is yet another choice of producing a constrained dynamics involving equal number of equations and unknowns in the constrained problem. Of note here is also the work of \cite{fan2005reflections} which proposes a generalization of the Gauss principle of Least Constraint. It is perhaps important to observe that Gauss' Least constraint principle, given full knowledge of the arrays $(f, n, w)$,
\begin{equation*}
\begin{aligned}
    & \min_b (b_i - f_i) (b_i - f_i)\\
    & \mbox{subject to } n^\alpha_i b_i + w^\alpha = 0,
\end{aligned}
\end{equation*}
is an additional, eminently physical, assumption. When a Lagrange multiplier formulation is employed to solve the optimization problem, the critical point is characterized by solutions of 
\[
b_i = f_i - \lambda^\alpha n^\alpha_i,
\]
identifying, in the process, the nature of the constraint force specific to this protocol. Moreover, when utilizing the Lagrange multiplier formulation of the problem, the uniqueness of the (instantaneous) values of the Lagrange multipliers rests on the invertibility of the matrix $n^\alpha_i n^\beta_i$ in
\[
n^\alpha_i n^\beta_i \lambda^\beta = w^\alpha + n^\alpha_i f_i,
\]
and this makes it clear that even with the Lagrange multiplier formulation of Gauss' principle, the constraint forces may not be unique and further conditions may then be required to induce even local-in-time uniqueness of the solutions $t \mapsto x(t)$ of the system \eqref{eq:constrained_disc} or its anholonomic counterpart.

Ad-hoc as the above assumptions on the form of the constraint force may seem, it is also extremely important to note that some assumption(s) beyond simply invoking the presence of a constraint force in the equations of motion and requiring that it expend no power when the system experiences internal kinematic constraints is necessary on physical grounds. As a simple example, consider the free (without any external applied force) motion of a single particle constrained to move on the plane $x_3 = 0$. If the only requirement of the constraint forces is that they maintain the constraint on the motion of the particle and expend no power, circular, in-plane motions would clearly be allowed, with the (in-plane component of the) constraint force pointing from the particle to the center of the circle - this is physical nonsense, of course.

In what follows, we demonstrate a family of variational principles for the equations of constrained analytical dynamics - dissipative or conservative - without ad-hoc assumptions, leaving enough flexibility for the use of specific models as described above, as well as for the general mathematical study of solutions to the equations, including that of the physically relevant conditions needed for inducing uniqueness of solutions for the initial value problem (ivp) of Newtonian classical mechanics of mass points.

\section{Dual variational principle for constrained particle dynamics}\label{sec:dual_vp}
None of the arguments and assumptions in Sec.~\ref{sec:intro_constr} shed light on when the final set of (un)constrained equations of motion can be expected to arise as the Euler-Lagrange equations of an action principle defined on trajectories/paths of an appropriately defined set of variables. Indeed, even in the unconstrained case, if the applied forces, $f$, were not assumed to arise as the gradient of a potential function in $x$, then it is generally believed that an action principle does not exist for the dynamics.

With the above considerations in mind, we would now like to propose a scheme that generates a family of variational principles for each of the forms of (un)constrained dynamics considered above. In fact, we will also allow for constraints that can do non-negative work along trajectories. As a natural result, the Euler-Lagrange equations of our formulation always involve the same number of dual variables as the number of primal equations (\eqref{eq:constrained_disc_1} and constraints), even when the primal system is formally underdetermined, as discussed earlier in this Section. Furthermore, unlike Hamilton's Principle that has the slightly unpleasant feature of requiring information on final-time boundary conditions on the primal position and velocities (also see discussion in \cite{galley2013classical}), a specification inconsistent with solving the well-set ivp arising from Newton's Laws, our dual variational principle exactly recovers as its Euler-Lagrange equations and natural boundary conditions the Newtonian ivp. Interestingly, our methodology can associate a family of dual action functionals with the primal Euler-Lagrange equations arising from a Hamiltonian structure. Our work is a natural application of ideas presented in \cite{action_2, action_3, KA1}, the last of which shows a computation of (approximate) solutions to Euler's equations for the the dynamics of a rigid body constrained at a single point under no torque as well as a dissipative torque (among other PDE examples like the linear heat and transport equations). Among the nontrivial PDE examples solved by our duality-based method are nonlinear elasticity and the inviscid Burgers equation \cite{singh2024hidden,kouskiya2024inviscid}.

In this section we develop a general framework for systems that are subject to possibly non-holonomic time-dependent constraints and dissipative forces.  We demonstrate the power of the method by applying it to the case of the dissipative and nonlinear Lorenz system (\ref{eq:lorenz}), obtaining the dual Lorenz action explicitly (\ref{E:dualLLorenz}). The reader may also consult Sec.~\ref{sec:gen_Pars} for another example.

As before, we are considering a system with $N$ constituent components, such as particles, with the $A$-th component having coordinates $(x_{1A},\ldots, x_{dA})\in {\mathbb R}^d$ and velocity denoted $(v_{1A},\ldots, v_{dA})$. The following constrained differential-algebraic system is considered, with $f, g, W$ being   suitably smooth given functions of their arguments:
\begin{equation}\label{eq:primal_sys}
    \begin{aligned}
        & \dot{x}_{iA}  = v_{iA}, \qquad  i = 1, \ldots, d; A = 1, \ldots, N \qquad \mbox{(position-velocity relationship)}\\
        & \dot{v}_{iA}  = f_{iA} (x, v, Q, R, t), \qquad \mbox{(Newton's Second Law)}\\
         & 0  = g^\alpha (x, v, t)  \qquad \alpha = 1, \ldots, m < d . N, \qquad  \mbox{(constraints)}\\
         & \dot{Q}_\Gamma  = R_\Gamma; \qquad \Gamma = 1,\ldots, \bar{\Gamma}, \quad m \leq \bar{\Gamma} \leq d . N \qquad  \mbox{(evolution of constraint parameters)}\\
         & 0 = W(x, v, Q, R, s, t), \qquad \mbox{(power expended by constraint forces)}\\
         & t \mapsto s(t) \in \mathbb{R} \ \mbox{is a slack variable to convert an inequality constraint into an equality}\\
       &  x^0_{iA} = x_{iA}(0) \qquad  \mbox{(initial condition on position)}\\
       &  v^0_{iA} = v_{iA}(0)  \qquad  \mbox{(initial condition on velocity)},
    \end{aligned}
\end{equation}
and it is required that the prescribed initial data $(x^0, v^0)$ be consistent with the constraints at the initial time, i.e.,  $0  = g(x^0, v^0, 0)$. The domain of the ${\mathbb R}^{dN}$-valued function $f=(f_{iA})$ is  an open subset of ${\mathbb R}^{dN}\times {\mathbb R}^{dN}\times {\mathbb R}^{{\bar{\Gamma}} }\times {\mathbb R}^{{{\bar{\Gamma}}}}\times [0,\infty)$, and $g$ is a smooth ${\mathbb R}^m$-valued function on ${\mathbb R}^{2dN+1}$. Further, $W$ is a smooth real-valued function on ${\mathbb R}^{2dN+2{\bar{\Gamma}}+2}$.
Solving the system (\ref{eq:primal_sys}) means finding a smooth mapping
$$t\mapsto \bigl(x(t), v(t), Q(t), R(t), s(t)\bigr)\in {\mathbb R}^{dN}\times {\mathbb R}^{dN}\times {\mathbb R}^{\bar{\Gamma}} \times {\mathbb R}^{\bar{\Gamma}} \times {\mathbb R},$$
with  $t\in [0,T]$, for some positive $T$, such that (\ref{eq:primal_sys}) holds when both sides are evaluated at each time instant $t$.

We assume the following structure for the forces
\begin{equation}\label{E:fiA}
f_{iA}(x, v, Q, R, t) = f^{(a)}_{iA}(x, v,t) - f^{(c)}_{iA}(x,v, Q, R, t),
\end{equation}
where $f^{(c)}$ represents the constraint force.

Examples of the constraint forces for the various models discussed are
\begin{subequations}
\label{eq:constr_types}
    \begin{align}
        f^{(c)}_{iA} & = Q^{(c)}_{iA} \qquad \mbox{no assumption made on the form of the constraint force}  \label{eq:constr_types_1}\\
        f^{(c)}_{iA} & = - Q_\alpha a^\alpha_{iA}(x, t) \qquad \mbox{d'Alembert; }  a \mbox{ is a known function of its arguments}\\
         f^{(c)}_{iA} & =  - Q_\alpha n^\alpha_{iA}(x, v,t) \qquad \mbox{Gauss; } n \mbox{ is a known function of its arguments}\\
          f^{(c)}_{iA} & = - \dot{Q}_\alpha a^\alpha_{iA} (x) - Q_\alpha \left( \frac{\partial a^\alpha_{iA}} {\partial x_{jB}}  - \frac{\partial a^\alpha_{jB}} {\partial x_{iA}} \right) v_{jB} \\
          & \qquad \mbox{Hamiltonian formalism for  a homogeneous, linear-in-velocity constraint}. \notag
    \end{align}
\end{subequations}
An example of constraint forces expending non-negative power, which also shows the use of a slack variable $s$, is
\begin{equation}\label{E:Wexample}
W(x, v, Q, R, s, t) := \sum_{i, a} m_A f^{(c)}_{iA} (x,v,Q,R,t)  v_{iA} - s^2 = 0.
\end{equation}
where $m_A$ is the mass of particle $A$; note that, by our notational convention, $f_{iA}$ is the $i$-th component of the force on the $A$-th particle divided by the mass $m_A$.

When $f^{(c)}(x,v,Q,R,t) = Q$ then the constraint parameter $Q$ has the meaning of the usual Lagrange multiplier. The form of the constraint force in the Hamiltonian formalism with constraints dictates that the constraint forces can depend on $\dot{Q}$; motivated by this, we allow for a general nonlinear dependence on $\dot{Q}$ of the constraint force, which requires, within our formalism, to introduce a new variable which we denote by $R := \dot{Q}$ and include it within our governing set of equations - this is no different, in principle, from the introduction of a new variable $v := \dot{x}$ in classical mechanics.

The totality of the \emph{primal} physical functions whose evolution is of interest are
\begin{equation}\label{E:defUxvQ}
U := (x,v,Q,R,s);
\end{equation}
the components of $U$ can be thought of as coordinate functions on an open subset of ${\mathbb R}^{2dN+2{\bar{\Gamma}}+1}$, and we are interested in motions $t\mapsto U(t)$ that satisfy the conditions (\ref{eq:primal_sys}). 

We introduce an array of \emph{dual} functions of time 
\begin{equation}\label{E:defarrayD}
\begin{aligned}
     & D := (\rho, \lambda, \mu, \Lambda, \kappa) \\
     & \rho_{iA}, \qquad i = 1,\ldots,d; A = 1, \ldots, N\\
     & \lambda_{iA}, \qquad i = 1,\ldots,d; A = 1, \ldots, N\\
     &  \mu_\alpha, \qquad \alpha = 1, \ldots, m < d . N\\
     & \kappa_\Gamma, \qquad \Gamma = 1,\ldots, \bar{\Gamma}, \quad m \leq \bar{\Gamma} \leq d . N
\end{aligned}
\end{equation}
to be used subsequently, as well as the array 
\begin{equation}\label{E:calD}
\dee := (\dot{\rho}, \rho, \dot{\lambda}, \lambda, \mu, \Lambda, \dot{\kappa}, \kappa).
\end{equation}
Technically, the components of $D$ can be thought of as coordinate functions on ${\mathbb R}^{2dN+m+1+{\bar{\Gamma}}}$.
Then define the pre-dual functional
\begin{equation}\label{E:Shat}
    \begin{aligned}
        \widehat{S}_H [x,v,Q,\rho, \lambda, \mu, \Lambda, \kappa] = - \rho(0) \cdot x^0 - \lambda(0) \cdot v^0 + \int_0^T \scl_H (U(t), \dee(t), t) \, dt,
    \end{aligned}
\end{equation}
where on the left the argument $(x,v,Q,\rho, \lambda, \mu, \Lambda, \kappa)$ is any smooth function   $[0,T]\to {\mathbb R}^{ 2dN+{\bar{\Gamma}}+2dN+m+{\bar{\Gamma}}}$, and the integrand on the right is
\begin{equation}\label{E:LH}
\scl_H (U, \dee, t) : = - \dot{\rho} \cdot x - \rho \cdot v - \dot{\lambda} \cdot v - \lambda \cdot f + \mu \cdot g + \Lambda W - \dot{\kappa} \cdot Q - \kappa \cdot R + H(U, \overline{U}(t)),
\end{equation}
and
\begin{equation}\label{E:Ubar}
\overline{U}(t) := (\bar{x}(t), \bar{v}(t), \bar{Q}(t), \bar{R}(t), \bar{s}(t))
\end{equation}
is a collection of the arbitrarily specified functions of time displayed, \emph{and $H$ is any smooth function of its arguments  that enables the construction of a function
\[
U^{(H)}(\dee, t) = \left( x^{(H)}(\dee, t), v^{(H)}(\dee, t), Q^{(H)}(\dee, t), R^{(H)}(\dee, t), s^{(H)}(\dee, t) \right)
\]
such that
\begin{equation}\label{eq:H_cond}
    \frac{\partial \scl_H}{\partial U} \left( U^{(H)} (\dee, t), \dee, t \right) = 0 \qquad \forall (\dee, t),
\end{equation}
i.e., $\frac{\partial \scl_H}{\partial U} (U, \dee, t) = 0$ should be solvable for $U$ in terms of $(\dee, t)$.} To be more precise, here $H$ is a real-valued smooth function on $({\mathbb R}^{2dN+2{\bar{\Gamma}}+1})^2$. (Note that $H$ does not refer to a Hamiltonian.)

\emph{We refer to the function $U^{(H)}$ as the dual-to-primal (DtP) mapping.}

The functional $\widehat{S}_H$ is obtained by taking appropriate scalar products of \eqref{eq:primal_sys} with the dual fields (Lagrange multipliers) $D$, integrating by parts over the time interval $[0,T]$, and applying the primal boundary conditions available and simply ignoring the remaining boundary terms at this point.

Then define the \emph{dual} functional
\begin{equation}\label{E:Shdual}
    \begin{aligned}
       &  S_H[\rho, \lambda, \mu, \Lambda, \kappa] = - \rho(0) \cdot x^0 - \lambda(0) \cdot v^0 + \int_0^T \scl_H \left(U^{(H)}(\dee(t), t), \dee(t), t \right) \, dt\\
       & \mbox{with Dirichlet boundary conditions} \quad \lambda(T), \rho(T), \kappa(T), \kappa(0) \quad \mbox{specified arbitrarily.}
    \end{aligned}
\end{equation}
Note that, on the left, $[\rho, \lambda, \mu, \Lambda, \kappa]$ is any smooth function   $[0,T]\to {\mathbb R}^{2dN+ m+1+{\bar{\Gamma}}  }$.

Utilizing the requirement \eqref{eq:H_cond}, and the boundary conditions
\[
\delta \lambda(T) = 0, \delta \rho(T) = 0, \delta \kappa(T) = 0, \delta \kappa(0) = 0,
\]
the first variation of $S_H$ is given by (using the short-hand and slight abuse of notation $U^{(H)}(t) = U^{(H)}(\dee(t), t)$ and $f^{(H)}(t) = f(U^{(H)}(t))$, $g^{(H)}(t) = g(U^{(H)}(t))$, $W^{(H)}(t) = W(U^{(H)}(t))$ )
\begin{equation}\label{E:delSH}
    \begin{aligned}
        \delta S_H & = - \delta \rho(0) \cdot x^0 - \delta \lambda(0) \cdot v^0 + \int_0^T \frac{\partial \scl_H}{\partial \dee} \left( U^{(H)} (\dee(t), t), \dee(t), t \right) \cdot \delta \dee \, dt \\
        & = \left(x^{(H)} (0) - x^0 \right) \cdot \delta \rho(0) + \int_0^T \delta \rho(t) \cdot \left( \dot{\overline{x^{(H)}}}(t) - v^{(H)}(t) \right) \, dt \\
        & \quad + \left(v^{(H)} (0) - v^0 \right) \cdot \delta \rho(0) + \int_0^T \delta \lambda(t) \cdot \left( \dot{\overline{v^{(H)}}}(t) - f^{(H)}(t) \right) \, dt \\
        & \quad +  \int^T_0 \delta \mu(t) \cdot  g^{(H)}(t) \, dt + \int^T_0 \delta \Lambda(t) W^{(H)}(t) \, dt +  \int^T_0 \delta \kappa(t) \cdot \left( \dot{\overline{Q^{(H)}}}(t) - R^{(H)}(t) \right) \, dt. 
    \end{aligned}
\end{equation}
Apart from a direct computation, this may also be understood by observing that $\scl_H(U,\dee,t)$ is affine in $\dee$ by construction. We note that the Euler-Lagrange equations of the dual functional $S_H$, \emph{parametrized by the function $H$}, is the set of primal equations \eqref{eq:primal_sys} with the replacement
\[
(U, f, g, W)  \rightarrow \left( U^{(H)}, f^{(H)}, g^{(H)}, W^{(H)} \right).
\]

To understand some salient properties of our variational principle, let us now consider a special case where no assumptions are made on the form of the constraint forces \eqref{eq:constr_types_1} so that
\[
f(x,v,Q,t) = f^{(a)}(x,v,t) - Q.
\]

For simplicity in conveying ideas, let us also assume that the power of constraint forces statement is presented as an equality so that the slack variable $s$ is not required, and that the functions $W$ is independent of $R$. Furthermore, we assume that the constraint function $g$ does not depend on $Q$ (and $s$). 

Then it suffices to choose the function $H$ to be a `shifted quadratic' form given by
\begin{equation}\label{eq:quadratic}
    H(x,v,Q,t) = \frac{1}{2} \left( c_x |x - \bar{x}(t)|^2 + c_v |v - \bar{v}(t)|^2 + c_Q |Q - \bar{Q}(t)|^2 \right),
\end{equation}
where $(x, v, Q)$ runs over  ${\mathbb R}^{3dN} $ here.

The key requirement \eqref{eq:H_cond} now becomes the condition that the following algebraic system of equations 
\begin{equation}\label{E:prLHprprimal}
    \begin{aligned}
        \frac{\partial \scl_H}{\partial x_{iA}}&: \qquad c_x (x - \bar{x})_{iA} - \dot{\rho}_{iA}  - \lambda_{jB} \frac{\p f^{(A)}_{jB}}{\p x_{iA}}(x,v,t) + \mu_\alpha \frac{\p g^\alpha}{\p x_{iA}}(x,v,t) + \Lambda \frac{\p W}{\p x_{iA}}(x,v,t) = 0 \\
        \frac{\partial \scl_H}{\partial v_{iA}}&: \qquad c_v (v - \bar{v})_{iA} - \rho_{iA} - \dot{\lambda}_{iA}  - \lambda_{jB} \frac{\p f^{(A)}_{jB}}{\p v_{iA}}(x,v,t) + \mu_\alpha \frac{\p g^\alpha}{\p v_{iA}}(x,v,t) + \Lambda \frac{\p W}{\p v_{iA}}(x,v,t) = 0 \\
        \frac{\partial \scl_H}{\partial Q_{iA}}&: \qquad c_Q (Q - \bar{Q})_{iA} + \lambda_{iA}  + \Lambda \frac{\p W}{\p Q_{iA}}(x,v,Q,t) = 0 \\
    \end{aligned}
\end{equation}
be solvable for $(x, v, Q)$ in terms of 
\begin{equation}\label{E:Dualt}
\dee = (\dot{\rho}, \rho, \dot{\lambda}, \lambda,\mu,\Lambda) \qquad \mbox{and} \qquad (\bar{x}(t), \bar{v}(t), \bar{Q}(t))
\end{equation}
to define the functions
\begin{equation}\label{E:xHvH}
    \begin{aligned}
         (\dee, t) & \mapsto x^{(H)}(\dee, t)\\
         (\dee, t) & \mapsto v^{(H)}(\dee, t)\\
         (\dee, t) & \mapsto Q^{(H)}(\dee, t).
    \end{aligned}
\end{equation}
Suppose, now, this can be done (and definitely in a perturbative sense, by choosing $c_x, c_v, c_q \to +\infty$). Then the following remarks are in order:
\begin{enumerate}
    \item Let ($t \mapsto \bar{x}(t)$, $t \mapsto \bar{v}(t)$, $t \mapsto \bar{Q}(t)$) be a solution to the primal problem \eqref{eq:primal_sys} in this special case for given initial data. Then, for the quadratic $H$ defined in \eqref{eq:quadratic} by these specified $\bar{(\cdot)}$ functions, \emph{the dual variational principle has a critical point, explicitly given by}
    \[
    \left( t \mapsto \rho(t) = 0 \in \mathbb{R}^{d.N}, \quad  t \mapsto \lambda(t) = 0 \in \mathbb{R}^{d.N}, \quad t \mapsto \mu(t) = 0 \in \mathbb{R}^{m}, \quad t \mapsto \Lambda(t) = 0 \in \mathbb{R} \right).
    \]

    This is an existence results for our proposed scheme. More importantly, it makes it reasonable to expect solutions to the dual problem to exist when the functions $(\bar{x}, \bar{v}, \bar{Q})$ chosen to define $H$ are close to actual solutions. Of course, even in other situations vastly different from these choices, solutions very likely exist, as motivated in \cite{KA1} through the computation of controlled approximations. Rigorous existence results for very weak and weak solutions in more general PDE contexts exist, e.g., the Euler equations and inviscid Burgers equation, see \cite{brenier2018initial,singh2024hidden}.
    \item We note that if the primal problem has a unique solution for given initial data and the variational dual problem has solutions for a collection of $H$'s and Dirichlet b.c.s, then the primal solutions defined by the differing class of $U^{(H)}$ are one and the same, regardless of the choice of $H$  (and b.c.s) defining them. This may be considered as a special type of gauge invariance of our procedure.
    \item For $c_x, c_v, c_Q$ large (and definitely when the primal system is linear),
    \begin{equation}\label{eq:bvp}
        x^{(H)} \sim \bar{x} + \frac{1}{c_x} \dot{\rho}, \qquad v^{(H)} \sim \bar{v} + \frac{1}{c_v} \dot{\lambda}
    \end{equation} 
and the dual Euler-Lagrange equation for $x^{(H)}, v^{(H)}$ become second-order ODE and the scheme proposes to solve it as a boundary value problem by arbitrarily specifying final-time conditions on ($\rho(T), \lambda(T)$). It is then reasonable to ask whether such a specification can act as an obstruction to describing the correct solution of the primal \emph{initial value problem} at time $T$, for solutions defined through the DtP mapping. The answer to the question is that such an obstruction does not arise as the mapping \eqref{eq:bvp} shows that $(x^{(H)}(T), v^{(H)}(T))$ are functions of $(\dot{\rho}(T), \dot{\lambda}(T))$, respectively, and specifying the values of $(\rho(T), \lambda(T))$ leaves the derivatives adjustable to the demands of solving the primal ivp.

\item Defining a function
\[
\mathbb{L}(D, \dot{D},t) := \scl_H \left( U^{(H)}(\dee,t), \dee, t\right)
\]
(since $\dee$ is a function of $(D,\dot{D},t)$), if 
\[
\frac{\p \mathbb{L}}{\p \dot{D}} (D, \dot{D}, t) = P
\]
is \emph{uniquely} solvable for 
\[
\dot{D} = \mathcal{R}(D, P, t),
\]
then it is an elementary result of Hamiltonian mechanics that a Hamiltonian
\[
\mathbb{H}(D,P,t) := P \cdot \mathcal{R}(D,P,t) - \mathbb{L}(D, \mathcal{R}(D, P,t), t)
\]
can be constructed by a Legendre transform as shown, and the corresponding Hamiltonian dynamics is essentially equivalent to our dual Lagrangian dynamics, i.e.,
\[
\frac{d D}{dt} = \frac{\p \mathbb{H}}{\p P} \,; \quad \frac{d P}{dt} = - \frac{\p \mathbb{H}}{\p D} \qquad \mbox{`}\Longleftrightarrow\mbox{'} \qquad \frac{d}{dt} \left( \frac{\p \mathbb{L}}{\p \dot{D}}\right) - \frac{\p \mathbb{L}}{\p D} = 0.
\]
Moreover, when $\mathbb{L}$ is independent of $t$ (certainly when the $H$ chosen to define $\scl_H$ is independent of $t$, e.g., by not involving any of the overhead $\ \bar{} \ $ functions), then
\[
\mathbb{H}(D, P)
\]
is a constant of the dual evolution and, hence, of the primal, possibly dissipative, evolution - since every dual evolution satisfying the dual E-L equation defines a primal motion by our basic scheme.

Since our scheme allows the association of, potentially, entire families of dual evolutions with a single primal system depending on the choice of the potential $H$, this suggests that there is the possibility of associating many constants of motion, in this sense, with a primal evolution, conservative or dissipative. This is an intriguing prospect whose implication needs to be explored in future work.

\item The dual action generated above for a dissipative system by our scheme and, in general, for a system of ODE of the type
\[
\dot{x} = F(x),
\]
including the particle systems that are the main subject of this paper, can be useful for computing (approximate) periodic orbits by looking for critical points of the functionals in the class of periodic functions (cf. \cite{rabinowitz}), complementary to the `Least-Squares' approaches in \cite{boghosian2011new, lan2004variational}. This is a potential application of our methods that is of considerable practical importance. 

For $T$ (a possible period of a periodic orbit to be determined), introducing the change of the time variable
\[
s = \frac{2 \pi t}{T}; \qquad \mathcal{P} := \frac{T}{2 \pi}
\]
one seeks to find $2 \pi$-periodic solutions $(x, \mathcal{P})$ of the system
\begin{equation*}
\frac{d x}{ds} = \mathcal{P} F(x); \qquad \frac{d \mathcal{P}}{ds} = 0.
\end{equation*}
The problem can now be approached as looking for critical points of dual functionals developed by our scheme on the fixed interval $[0,2\pi]$ within the class of functions satisfying periodic boundary conditions on the dual fields. The flexibility provided by our scheme through the choice of the function $H$ which, in turn, allows the further use of the `base states' $\bar{x}$ may be expected to benefit the search for identifying a substantial class of periodic orbits.
\end{enumerate}

The proposed methodology is general and applies to a large collection of ODE systems (converted to first-order form). To explicitly demonstrate the `algorithmic' nature of the scheme, we apply our method to the dissipative and nonlinear Lorenz system \cite{lorenz1963deterministic} written as (with $A,R,B > 0$ as given parameters)
\begin{equation}\label{eq:lorenz}
\begin{aligned}
    & \dot{x} - A(y - x) = 0; \qquad \dot{y} - x(R - z) + y = 0; \qquad \dot{z} - xy + Bz = 0\\
    & \mbox{with initial conditions } \quad x(0) = x^0; \quad y(0) = y^0; \quad z(0) = z^0.
\end{aligned} 
\end{equation}
It is known that the Lorenz dynamics is dissipative (and chaotic for a certain parameter range) and long-time solutions are restricted to a small neighborhood of a bounded attracting set of phase space after, possibly, an initial transient; we will assume that initial conditions are chosen within a bounded region of phase space, which contains the aforementioned attracting set.

Next, with the choice 
\[
H(x,y,z, t) = \frac{1}{2} c \left( (x - \bar{x}(t))^2 + (y - \bar{y}(t))^2 + (z - \bar{z}(t))^2\right),
\]
where the functions $(\bar{x}, \bar{y}, \bar{z})$ of time are free to choose as is the constant $c >0$, form the pre-dual functional
\begin{equation*}
    \begin{aligned}
        \widehat{S}_H[U, D] & = \int_0^T \scl_H \left(U,\dee \right) \, dt - \lambda(0) x^0 - \mu(0) y^0 - \gamma(0) z^0 \mbox{ where }\\
         \scl_H \left(U, \dee \right) & = - x \dot{\lambda} - \lambda A (y -x) - y \dot{\mu} - \mu x (R - z) + \mu y - z \dot{\gamma} - \gamma xy + \gamma B z + H(U,t),
    \end{aligned}
\end{equation*}
where
\begin{equation*}
     U = (x,y,z); \qquad D = (\lambda, \mu, \gamma); \qquad \dee = (\lambda, \dot{\lambda}, \mu, \dot{\mu}, \gamma, \dot{\gamma}).
\end{equation*}

The DtP mapping $U^{(H)}(\dee)$ is generated from solving the equations
\begin{equation*}
\begin{rcases}
    \begin{aligned}
        \frac{\p \scl_H}{\p x} = 0&: \qquad - \dot{\lambda} + A \lambda - \mu R - \gamma y + \mu z + c(x - \bar{x})  = 0\\
        \frac{\p \scl_H}{\p y} = 0&: \qquad  - \lambda A - \dot{\mu} + \mu - \gamma x + c( y - \bar{y}) = 0 \\
        \frac{\p \scl_H}{\p z} = 0&: \qquad \mu x - \dot{\gamma} + \gamma B + c(z - \bar{z}) = 0
    \end{aligned}
    \end{rcases}
    \qquad \mathbb{A}\big|_D \left( U^{(H)} - \bar{U} \right) =  p\big|_{(\dee, \bar{U})}
\end{equation*}
with the matrix defined as
\begin{equation*}
\mathbb{A}\big|_D = c
\begin{bmatrix}
    1 & - \frac{\gamma}{c} & \frac{\mu}{c} \\
    - \frac{\gamma}{c} & 1 & 0 \\
    \frac{\mu}{c} & 0 & 1
\end{bmatrix}
\quad \mbox{ with } \quad \left(\mathbb{A}\big|_D \right)^{-1} = \frac{1}{c \left( 1 - \frac{\gamma^2}{c^2} - \frac{\mu^2}{c^2}\right)}
\begin{bmatrix}
1 & \frac{\gamma}{c} & - \frac{\mu}{c}\\
\frac{\gamma}{c} & \left(1 - \frac{\mu^2}{c^2}\right) & - \frac{\gamma \mu}{c^2} \\
- \frac{\mu}{c} & -\frac{\gamma \mu}{c^2} & \left( 1 - \frac{\gamma^2}{c^2} \right)
\end{bmatrix}
=: \mathbb{B}\big|_D,
\end{equation*}
and the vectors as
\begin{equation*}
U^{(H)} - \bar{U} = 
\begin{bmatrix} 
x^{(H)} - \bar{x} \\  y^{(H)} - \bar{y} \\ z^{(H)} - \bar{z} 
\end{bmatrix}; \qquad 
    p\big|_{(\dee, \bar{U})} = \begin{bmatrix}
    \gamma \bar{y} - \mu \bar{z} + \dot{\lambda} - A \lambda + R \mu \\
    \gamma \bar{x} + \dot{\mu} - \mu + A \lambda \\
    - \mu \bar{x} + \dot{\gamma} - B \gamma
\end{bmatrix},
\end{equation*}
with solution given by
\begin{equation}\label{eq:lorenz_soln}
U^{(H)} - \bar{U} = \mathbb{B}\big|_D \ p\big|_{(\dee, \bar{U})}.
\end{equation}
The inverse matrix exists by making a choice of $c$ large enough (using the boundedness of the flow).
Computation, with suitable algebraic grouping, shows that
\begin{equation*}
    \begin{aligned}
        \widehat{S}_H[ U^{(H)}, D] & = \int_0^T  - ( x^{(H)} - \bar{x}) \left(  \dot{\lambda} - \lambda A + R \mu + \gamma \bar{y} - \mu \bar{z}   \right) \, dt - \lambda(0)x^0\\
         & \qquad + \int_0^T - (y^{(H)} - \bar{y}) \left( \lambda A + \dot{\mu} - \mu  + \gamma \bar{x} \right) \, dt  - \mu(0) y^0\\
         & \qquad + \int_0^T - (z^{(H)} - \bar{z}) \left( \dot{\gamma} + B \gamma - \mu \bar{x} \right) \, dt - \gamma(0) z^0\\
         & \qquad - \int_0^T \bar{U} \cdot p\big|_{(\dee,\bar{U})} \, dt + \int_0^T \mu \left(x^{(H)} - \bar{x}\right) \left(z^{(H)} - \bar{z} \right) - \gamma \left(x^{(H)} - \bar{x}\right) \left(y^{(H)} - \bar{y} \right)\, dt \\
         & \qquad + \int_0^T - \mu \bar{x} \bar{z} + \gamma \bar{x} \bar{y} + H\left(U^{(H)}, t \right) \, dt.
    \end{aligned}
\end{equation*}
Noting, further, that
\begin{equation*}
    \begin{aligned}
       \frac{1}{2} (U - \bar{U}) \cdot \mathbb{A} ( U - \bar{U}) & = \frac{1}{2} c\left( \left(x - \bar{x} \right)^2 + \left(y - \bar{y} \right)^2 + \left(z - \bar{z} \right)^2 \right) \\
       & \quad  + \mu \left(x - \bar{x}\right) \left(z - \bar{z} \right) - \gamma \left(x - \bar{x}\right) \left(y - \bar{y} \right),
    \end{aligned}
\end{equation*}
we obtain the corresponding \emph{dual Lorenz action} as:
\begin{equation}\label{E:dualLLorenz}
\begin{aligned}
 S_H[D] & =  \int_0^T \left( - \frac{1}{2} p\big|_{(\dee, \bar{U})} \cdot \mathbb{B}\big|_D  \  p \big|_{(\dee, \bar{U})} - \bar{U} \cdot p \big|_{(\dee, \bar{U})} - \mu \bar{x} \bar{z} + \gamma \bar{x}\bar{y} \right) \, dt \\
   & \quad  - \lambda(0) x^0 - \mu(0) y^0 - \gamma(0) z^0  \\
   & \mbox{ with } \lambda(T), \mu(T), \gamma(T) \mbox{ specified arbitrarily}.
\end{aligned}
\end{equation}
Our formalism and results guarantee  that the corresponding E-L equation and natural boundary conditions lead to
\begin{equation*}
\begin{aligned}
    & \frac{d{x}^{(H)}}{dt} - A\left(y^{(H)} - x^{(H)}\right) = 0; \quad \frac{d{y}^{(H)}}{dt}- x^{(H)} \left(R - z^{(H)} \right) + y^{(H)} = 0; \quad \frac{d{z}^{(H)}}{dt} - x^{(H)}y^{(H)} + Bz^{(H)} = 0\\
    &  \quad x^{(H)}\big|_{t = 0} = x^0; \quad y^{(H)}\big|_{t = 0} = y^0; \quad z^{(H)}\big|_{t = 0} = z^0.
\end{aligned} 
\end{equation*}
We note from \eqref{eq:lorenz_soln} that if $(t \mapsto \bar{x}(t), t \mapsto \bar{y}(t), t \mapsto \bar{z}(t))$ is a solution of the Lorenz system then the corresponding dual Lorenz action has a critical point given by $(t \mapsto \lambda(t) = 0, t \mapsto \mu(t) = 0, t \mapsto \gamma(t) = 0)$. In passing, we also note that for $(\bar{x} = 0, \bar{y} = 0, \bar{z} = 0)$ the dual Lagrangian is a negative semi-definite form for $c \gg 1$ (assuming boundedness of the dual flow).

Based on what has been said in remark 4.~above, it can be checked that when there is a solution to the dual Lorenz problem, defining (with $D = (\lambda, \mu, \gamma)$)
\begin{equation*}
    \begin{aligned}
        \mathbb{L}(D, \dot{D},t) & = - \frac{1}{2} p\big|_{(D, \dot{D}, \bar{U}(t))} \cdot \mathbb{B}\big|_D  \  p \big|_{(D, \dot{D}, \bar{U}(t))} - \bar{U}\big|_t \cdot p \big|_{(D, \dot{D}, \bar{U}(t))} - \mu (\bar{x} \bar{z})\big|_t + \gamma (\bar{x}\bar{y})\big|_t \\
        \tilde{p}\big|_{(D, \bar{U})} & := p\big|_{(D, \dot{D}, \bar{U})} - \dot{D}\\
        \mathcal{R}(D, P, t) & = - \mathbb{A}\big|_D \left( P + \bar{U}(t) \right) - \tilde{p} \big|_{(D, \bar{U}(t))},
    \end{aligned}
\end{equation*}
a Hamiltonian
\[
\mathbb{H}(D,P) := \mathcal{R}(D, P, t) \cdot P - \mathbb{L}(D, \mathcal{R}(D,P,t), t)
\]
can be associated with the dissipative Lorenz evolution, and it is a constant of the Lorenz motion when $\bar{U}$ is constant (in the sense of the remark 4).
\section{Generalization of the example of Pars \cite{pars1954variational}}\label{sec:gen_Pars}
We consider the system (with all particle masses assumed equal to $1$ in appropriate units)
\begin{subequations}\label{eq:ex_sys}
    \begin{align}
         \dot{x}_i - v_i &  = 0; \qquad  i = 1,\dots, M \in \mathbb{Z}^+  \label{eq:ex_sys_1}\\
         \dot{v}_i + Q_i & = 0 \label{eq:ex_sys_2}\\
         (b_i + L_{ij} x_j) v_i  & = 0 \label{eq:ex_sys_3}\\
         Q_i v_i - \frac{1}{2} s^2 & = 0 \label{eq:ex_sys_4}\\
        \mbox{with initial conditions} &  \notag\\
         x_i(0) &  = x^0_i  \label{eq:ex_sys_5}\\
         v_i(0)  & = v^0_i \label{eq:ex_sys_6} \\
          \left(b_i + L_{ij}x^0_j \right) v^0_i  & = 0.  \label{eq:ex_sys_7}
    \end{align}
\end{subequations}
Here $L,b$ are a given constant matrix and a vector, respectively, and the initial data $(x^0, v^0)$ is assumed to be consistent with \eqref{eq:ex_sys_7}. The system above can represent holonomic, anholonomic, and non-integrable constraints, including the example of Pars \cite{pars1954variational,cronstrom2007existence}.

It should be noted that (\ref{eq:ex_sys_4}, \ref{eq:ex_sys_2}) imply that along any solution of \eqref{eq:ex_sys} the rate of change of kinetic energy is non-positive, i.e., 
\begin{equation}\label{E:Kdot}
\dot{K} \leq 0,\qquad K = \frac{1}{2} v_i v_i. 
\end{equation}
In case external forces and forces of interaction arising from a potential $E(x)$ were to be involved to give the equation of motion
\[
\dot{v} + Q = f^{ext} - \p_x E,
\] 
the corresponding power of constraints statement would be an expression of non-negative mechanical dissipation given by 
\[
Q \cdot v = f^{ext} \cdot v - \frac{d}{dt} (E + K) \geq 0.
\] 

Denoting
\[
U = (x,v,Q,s); \qquad D = (\rho, \lambda, \mu, \Lambda); \qquad \dee = (\dot{\rho}, \rho, \dot{\lambda}, \lambda, \mu, \Lambda),
\]
we have
\[
\widehat{S}_H[x,v,Q,s,\rho, \lambda, \mu, \Lambda] = \int^T_0 \scl_H(U(t), \dee(t), t) \, dt - \rho(0) \cdot x^0 - \lambda(0) \cdot v^0,
\]
where
\begin{equation}\label{eq:gen_pars_lagrangian}
    \begin{aligned}
       \scl_H(U, \dee, t) & = - x \cdot \dot{\rho} - v \cdot \rho - v \cdot \dot{\lambda}  + \lambda \cdot Q + \mu v \cdot (L x) + \mu b \cdot v + \Lambda Q \cdot v - \frac{1}{2} \Lambda s^2 \\
       & \quad + \frac{1}{2} \left( c_x |x - \bar{x}(t)|^2 + c_v |v - \bar{v}(t)|^2 + c_Q |Q - \bar{Q}(t)|^2 + c_s (s - \bar{s}(t))^2 \right).
    \end{aligned}
\end{equation}
The DtP mapping is obtained from solving the system
\begin{equation}\label{eq:ex_dtp}
    \begin{aligned}
        \frac{\p \scl_H}{\p x_i} &: \qquad - \dot{\rho}_i + \mu L_{ki} v_k + c_x (x_i - \bar{x}_i) = 0 \\
        \frac{\p \scl_H}{\p v_i} &: \qquad - \rho_i - \dot{\lambda}_i + \mu b_i + \mu L_{ij} x_j + \Lambda Q_i + c_v (v_i - \bar{v}_i) = 0 \\
        \frac{\p \scl_H}{\p Q_i} &: \qquad \lambda_i + \Lambda v_i + c_Q (Q_i - \bar{Q}_i) = 0\\
        \frac{\p \scl_H}{\p s} &: \qquad  -\Lambda s + c_s (s - \bar{s}) = 0.
    \end{aligned}
\end{equation}


We now \emph{assume} that the functions $\mu^2$ and $\Lambda^2$ are bounded above in $[0,T]$ so that $c_x, c_v, c_Q$ can be chosen large enough to make the matrix
\[
\mathbb{M}_{ik} (\mu^2, \Lambda^2) := \delta_{ik} - \frac{1}{c_x c_v} \mu^2   L_{ij}L_{kj} - \frac{1}{c_Q c_v} \Lambda^2 \delta_{ik}
\]
invertible.

Then the solution of \eqref{eq:ex_dtp} is given by
\begin{equation}\label{eq:gen_pars_dtp}
    \begin{aligned}
        v_k^{(H)}(\dee, t) & = (\mathbb{M}^{-1})_{ki} \bar{v}_i(t) + \frac{1}{c_v}(\mathbb{M}^{-1})_{ki}\left( - \mu L_{ij} \bar{x}_j(t) - \Lambda \bar{Q}_i(t) + \rho_i + \dot{\lambda}_i - \mu b_i - \frac{1}{c_x}  \mu L_{ij} \dot{\rho}_j + \frac{1}{c_Q} \Lambda \lambda_i  \right)\\
        x^{(H)}_k(\dee, t) & = \bar{x}_i(t) + \frac{1}{c_x} \dot{\rho}_i - \frac{1}{c_x}  \mu L_{ki} v^{(H)}_k(\dee, t)\\
        Q^{(H)}_i (\dee, t)& = \bar{Q}_i(t) - \frac{1}{c_Q} \lambda_i - \frac{1}{c_Q} \Lambda v^{(H)}_i(\dee, t)\\
        s^{(H)} (\dee, t)& = \frac{ c_s \bar{s}(t)}{c_s - \Lambda}.
    \end{aligned}
\end{equation}
The dual functional is then given by
\begin{equation}\label{eq:gen_pars_dual_action}
 S_H[D] = \int^T_0 \scl_H \left(U^{(H)}(\dee(t),t), \dee(t), t \right) \, dt - \rho(0) \cdot x^0 - \lambda(0) \cdot v^0,   
\end{equation}
with
\[
\rho(T), \lambda(T) \quad \mbox{specified (arbitrarily)}.
\]
Since
\[
\frac{\p \scl_H}{\p U} \left( U^{(H)}(\dee, t), \dee, t \right) = 0 \qquad \mbox{for all } (\dee, t) \ \mbox{by design},
\]
we have
\begin{equation*}
    \begin{aligned}
        \delta S_H = \int^T_0  \frac{\p \scl_H}{\p \dee} \left( U^{(H)}(\dee, t), \dee, t \right) \cdot \delta \dee\, dt - \delta \rho(0) \cdot x^0 - \delta \lambda(0) \cdot v^0
    \end{aligned}
\end{equation*}
and repeating the general arguments of Sec.~\ref{sec:dual_vp} it can be checked that the primal system \eqref{eq:ex_sys} is recovered with the replacement 
$$(x(t), v(t), Q(t), s(t)) := \left(x^{(H)}(\dee(t),t), v^{(H)}(\dee(t),t), Q^{(H)}(\dee(t),t), s^{(H)}(\dee(t),t) \right).$$

\subsection{The example of Pars: variations on the theme}\label{sec:ex_pars}

A dynamical system is holonomic if its  evolution is specified by applied forces and constraints, possibly time-dependent, defined on the coordinates specifying the constituents of the system.  
If $\{x_i\}$ are coordinates specifying a state of the system,   holonomic dynamics is given by an equation for the accelerations of the form ${\ddot x}_i   = F_i^{(a)}(x,t)$ and constraints $g^\alpha(x,t)=0$, for $\alpha$ running over a finite set of indices,   and $F_i^{(a)}$ are the applied forces. 

Our interest is mainly in non-holonomic,   possibly dissipative, systems, specifically those where the constraints are of the form $g^\alpha(x,v, t)=0$ and forces of the form $F_i^{(a)}(x,v,t)$, where $v$ runs over the velocities of the constituents of the system.  For a brief history and a modern perspective, as well as other references, on non-holonomic dynamics we refer to \cite{BMZ2004}.

For holonomic systems, the traditional method is the d'Alembert principle which, in its basic form, states that the work done for virtual displacements by the constraint forces is zero  (the d'Alembert principle also applies to systems with constraints that are linear, but not affine, in the velocities); technically, this means that, with $F^{(a)}_i$ denoting the  $i$-th constituent component of the applied non-constraint force,
\begin{equation}\label{E:mixdotsxiF}
\sum_{i}(m_i{\ddot x}_i-F^{(a)}_i)\delta x_i=0
\end{equation}
for all variations $\{\delta x_i\}$ for which the constraints are maintained at each fixed time $t$;
\begin{equation}\label{E:galphadxi}
\sum_i \frac{\partial g^{\alpha}(x,t)}{\partial x_i}\delta x_i=0.
\end{equation}
In terms of Lagrange multipliers $\lambda_\alpha(t,x)$, one for each constraint $g_\alpha$, we have then
\begin{equation}\label{E:mixdotsxiFlambdai}
 m_i{\ddot x}_i-F^{(a)}_i-\lambda_\alpha \frac{\partial g^{\alpha}(x,t)}{\partial x_i}=0,
\end{equation}
with summation over the repeated index $\alpha$. Thus the solution $t\mapsto x(t)$ solves these equations and satisfies the constraints. In the case where the constraints are   \emph{time-independent}, the power expended by the constraint forces is then
\begin{equation}\label{E:mxdotslambda}
    (m_i{\ddot x}_i-F^{(a)}_i){\dot x}_i = \lambda_{\alpha}\frac{dg^{\alpha}\bigl(x(t)\bigr)}{dt},
\end{equation}
which is $0$ if the solution $t\mapsto x(t)$ satisfies the constraints.
Assuming that the applied force comes from a potential, the equation (\ref{E:mxdotslambda}) can be obtained as the Euler-Lagrange equations from an action of the form $\int L\bigl(t, x(t), {\dot x}(t),\lambda(t)\bigr)\,dt$.  For non-holonomic dynamics, with velocity-dependent constraints  (and/or velocity dependent applied forces) this classical method generally fails. As already discussed, our framework provides an action functional, explicitly involving the initial configuration of the system,  whose E-L equations coincide with the equations of motion. In this section we consider a non-holonomic system, with equations of motion given in (\ref{eq:pars}), and work through our framework  and obtain the  dual  action functional (\ref{E:SHPars}) for this system.

Pars \cite{pars1954variational} studied two examples of  non-holonomic systems in relation to variational principles.
One such system consists of the motion of a single particle in 3-d space without any applied forces but subject to the requirement that the path $t\mapsto x(t)$ must respect the constraint:

\[
x_3 d{x}_1 - d{x}_2 = 0.
\]
In particular, the velocity $\dot x$ and position coordinates $x$ must satisfy:

\[
x_3 \dot{x}_1 - \dot{x}_2 = 0.
\]
The differential form $\eta=x_3dx_1-dx_2$ gives a  non-holonomic constraint, in the sense that it cannot be expressed as $\mu\,dg$, for any constraint function $g$ and `intergrating factor' $\mu$, as any such form satisfies
$$(\mu\,dg)\wedge d(\mu\,dg) =0$$
whereas $\eta\wedge d\eta=-dx_2\wedge dx_3\wedge dx_1\neq 0$.

In the following, we will assume $C^2$ differentiability of all functions on the time interval $[0,T]$. Thus, the governing system of equations is
\begin{subequations}\label{eq:pars}
\begin{align}
    & \dot{x}_i - v_i = 0\\
    & \dot{v}_i + Q_i = 0 \label{eq:pars_linmom}\\
    & x_3 v_1 - v_2 = 0 \label{eq:pars_3}\\
    & x_i(0) = x_i^0; \qquad v_i(0) = v^0_i; \qquad x_3^0 v^0_1 - v^0_2  = 0; \quad \mbox{(initial data consistent with constraint)}.
\end{align}
\end{subequations}
The constraint can be written as $a_i v_i = 0$, where $a = (x_3, -1, 0)$, as well as 
\begin{equation}\label{eq:gauss_constr_form}
    n_i \ddot{x}_i + w = 0
\end{equation} 
with 
\[
n = a = (x_3, - 1, 0); \qquad w = \dot{x}_3 \dot{x}_1.
\]
In terms of the matrix $L$ and vector $b$, $L_{13} = 1, b_2 = -1$, respectively, with all other components of the arrays being $0$. Gauss' principle of Least Constraint can be used to reduce this system to a system of three equations in three unknowns whose solutions satisfy the constraint \eqref{eq:pars_3}, by \emph{assuming}
\[
Q_i := \lambda^* n_i = \lambda^* a_i
\]
with $\lambda^*$ evaluated from substituting $Q_i = \lambda^* n_i$ and \eqref{eq:pars_linmom}  into the constraint written as 
\[
0 = n_i \dot{v}_i + w  = - n_i Q_i + w = - \lambda^* n \cdot n + w = - \lambda^* x_3^2 - \lambda^* + \dot{x}_3 \dot{x}_1 \Longrightarrow \lambda^* = \frac{v_3 v_1}{( 1 + x_3^2)}.
\]
Then the reduced equations of motion satisfying the constraint are:
\begin{equation}\label{eq:pars_red_soln}
    \ddot{x}_1 = \frac{- \dot{x}_1 x_3 \dot{x}_3  }{(1 + x_3^2)}; \qquad \ddot{x}_2 = \frac{ \dot{x}_1  \dot{x}_3  }{(1 + x_3^2)}; \qquad \ddot{x}_3 = 0.
\end{equation}
Given the homogeneous, linear-in-velocity constraint $a \cdot v = 0$ and the choice of the form of the constraint force, it is clear that the constraint force of the formalism expends no power along solutions of \eqref{eq:pars}, i.e.,
\[
Q \cdot v = 0
\]
as well as all trajectories satisfying only the constraint \eqref{eq:pars_3}.

Suppose we now define a solution to \eqref{eq:pars_red_soln} as $ t \mapsto \bar{x}(t), t \mapsto \bar{v}(t)$ and the corresponding constraint force $t \mapsto \lambda^*(t)a(t) = \bar{Q}(t)$. For $M = 3$ in \eqref{eq:ex_sys}, defining $\scl_H$ in \eqref{eq:gen_pars_lagrangian} and the dual action in \eqref{eq:gen_pars_dual_action} using these barred functions with $t \mapsto \bar{s}(t) = 0$ and the functions $x^{(H)}, v^{(H)}, Q^{(H)}, s^{(H)}$ defined in \eqref{eq:gen_pars_dtp} replacing $x,v, Q, s$, we have the result that the Euler-Lagrange equations of the action \eqref{eq:gen_pars_dual_action} recover \eqref{eq:pars} (and $Q \cdot v = 0$) with the change of variables $x \to x^{(H)}, v \to v^{(H)}, Q \to, Q^{(H)}$ and that a solution to this system of equations, viewed as equations for determining the dual state, is given by $t \mapsto \rho(t) = 0, t \mapsto \lambda(t) = 0, t \mapsto \mu(t) =0, t \mapsto \Lambda(t) = 0$. In turn, $x^{(H)}, v^{(H)}, Q^{(H)}, s^{(H)}$ in \eqref{eq:gen_pars_dtp} defined in terms of this dual set of functions solve the Pars system \eqref{eq:pars}.

Let us now consider a constraint force of the form
\begin{equation}\label{eq:pars_damped_constr_force}
    Q = \lambda^* \, a(x) + a(x) \times v + \nu \, v; \qquad \nu \geq 0, \mbox{ a scalar constant}.
\end{equation}
Then, it is reasonable to expect solutions $(t \mapsto (x(t), v(t), Q(t))$ to the system \eqref{eq:pars} with this particular form of $Q$ by considering the system
\begin{subequations}\label{eq:pars_mod}
\begin{align}
    & \dot{x}_i - v_i = 0  \label{eq:pars_mod_1}\\
    & \dot{v}_i = - \lambda^* n_i(x) - \veps_{ijk} n_j(x) v_k - \nu v_i \label{eq:pars_mod_2}\\
    & x_3 v_1 - v_2 = 0  \label{eq:pars_mod_3}\\
    & ( \lambda^* n_i(x) + \veps_{ijk} n_j(x) v_k + \nu v_i) v_i - \frac{1}{2} s^2 = 0  \label{eq:pars_mod_4}\\
    & x_i(0) = p_i^0; \qquad v_i(0) = v^0_i; \qquad x_3(0) v^0_1 - v^0_2  = 0; \quad \mbox{(initial data consistent with constraint)},  \label{eq:pars_mod_5}
\end{align}
\end{subequations}
and obtaining a solution $(t \mapsto (x(t), v(t), \lambda^*(t))$ to the constrained ivp given by (\ref{eq:pars_mod_1}, \ref{eq:pars_mod_2}, \ref{eq:pars_mod_3}, \ref{eq:pars_mod_5}), as well as a solution to \eqref{eq:pars_mod} by using that solution to assign the value of $s^2$ from \eqref{eq:pars_mod_4}. 

Indeed, a solution can be generated as follows: with $n \times v$ given by
\[
n \times v = (- \dot{x}_3, - x_3 \dot{x}_3, x_3 v_2 + v_1),
\]
substitute \eqref{eq:pars_mod_2} (corresponding to the constraint force \eqref{eq:pars_damped_constr_force}) into \eqref{eq:gauss_constr_form} to solve for $\lambda^*$ as
\[
\lambda^*(x, v; \nu) = \frac{v_3 v_1 + \nu( v_2 - x_3 v_1)}{x_3^2 + 1}.
\]
With this expression of $\lambda^*(x, v; \nu)$, solve the system of ODE (\ref{eq:pars_mod_1}, \ref{eq:pars_mod_2}, \ref{eq:pars_mod_5}). It can be checked that, by construction, the obtained solution satisfies the constraint \eqref{eq:pars_mod_3}, since the initial data $\left(x^0, v^0 \right)$ is required to satisfy \eqref{eq:pars_mod_5}.

 Now, \eqref{eq:pars_mod_3} is simply the expression $n_i(x(t)) v_i(t) = 0$, so that \eqref{eq:pars_mod_4} becomes
\[
2 \nu \, v_i v_i = s^2
\]
and assigning $t \mapsto s(t)$ to be any of $\pm \sqrt{2 \nu \, v(t) \cdot v(t)}$ satisfies \eqref{eq:pars_mod_4}. 

For $\nu = 0$, the constraint force expends no power but does not satisfy the d'Alembert requirement of being in the direction of $a(x)$; moreover, the equations of motion are given by
\begin{equation}\label{eq:pars_ex_2}
    \ddot{x}_1 = \frac{- \dot{x}_1 x_3  \dot{x}_3 }{(1 + x_3^2)}  + \dot{x}_3; \qquad \ddot{x}_2 = \frac{ \dot{x}_1 \dot{x}_3  }{(1 + x_3^2)} + x_3 \dot{x}_3; \qquad \ddot{x}_3 = - x_3 \dot{x}_2 - \dot{x}_1.
\end{equation}
These equations cannot yield the same motion as \eqref{eq:pars_red_soln} from identical, but generic, initial conditions, even for short times (it suffices to note the evolution of $x_3$ for the two cases).

Of course, for all choices of the constraint force, our scheme presented in Sec.~\ref{sec:dual_vp}-\ref{sec:gen_Pars} generates an action principle for the corresponding system of governing constrained ODE. For example, consider \eqref{eq:pars_red_soln} and denote
\begin{equation}\label{eq:pars_defz}
\begin{aligned}
     &  x_3(t) = v^0_3 t + x_3^0 =: z(t); \qquad \dot{v}_1  - \frac{v_1 z v_3^0}{1+z^2} = 0; \qquad \dot{v}_2 - \frac{v_1 v_3^0}{1 + z^2} = 0,\\
     & v_1(0) = v_1^0; \qquad v_2(0) = v_2^0.
\end{aligned}
\end{equation}
Next, with the choice $H(v_1,v_2) = \frac{1}{2} \left( v_1^2 + v_2^2\right)$, form the pre-dual functional
\begin{equation*}
    \begin{aligned}
        \widehat{S}_H[v,\lambda] & = \int_0^T \scl_H \left(v, \lambda, \dot{\lambda} \right) \, dt - \lambda_1(0) v_1^0 - \lambda_2(0) v_2^0 \\
        \mbox{where } \scl_H \left(v, \lambda, \dot{\lambda} \right) & = - v_1 \dot{\lambda}_1 - \frac{ \lambda_1 v_1 z v_3^0}{1+z^2} - v_2 \dot{\lambda}_2 - \frac{\lambda_2 v_1 v_3^0}{1 + z^2} + \frac{1}{2} \left( v_1^2 + v_2^2\right).
    \end{aligned}
\end{equation*}
The DtP mapping $v^{(H)} \left(\lambda, \dot{\lambda} \right)$ is generated from solving the equations
\begin{equation*}
    \begin{aligned}
        \frac{\p \scl_H}{\p v_1} = 0&: \qquad v_1^{(H)}\left(\lambda, \dot{\lambda} \right) =   \dot{\lambda}_1 + \frac{ \lambda_1 z v_3^0}{1+z^2} + \frac{\lambda_2 v_3^0}{1 + z^2} \\
        \frac{\p \scl_H}{\p v_2} = 0&: \qquad  v_2^{(H)}\left(\lambda, \dot{\lambda} \right) =  \dot{\lambda}_2.
    \end{aligned}
\end{equation*}
Then the \emph{dual action} is given by $S_H[\lambda] = S_H[v^{(H)}[\lambda], \lambda]$:
 
\begin{equation}\label{E:SHPars}
    \begin{aligned}
        S_H[\lambda] & = - \frac{1}{2} \int_0^T \left( \left(v^{(H)}_1\right)^2  + \left(v^{(H)}_1\right)^2 \right) \, dt - \lambda_1(0) v_1^0 - \lambda_2(0) v_2^0 \\
        & = - \frac{1}{2} \int_0^T \left( \left( \dot{\lambda}_1 + \frac{ \lambda_1 z v_3^0}{1+z^2} + \frac{\lambda_2 v_3^0}{1 + z^2} \right)^2  + \left( \dot{\lambda}_2  \right)^2 \right) \, dt - \lambda_1(0) v_1^0 - \lambda_2(0) v_2^0,\\
        & \qquad \mbox{with } \lambda_1(T), \lambda_2(T) \mbox{ specified arbitrarily,}
    \end{aligned}
\end{equation}

along with the guarantee that its E-L equation and natural boundary conditions are
\[
\qquad \dot{\overline{{v}^{(H)}_1}}  - \frac{v^{(H)}_1 z v_3^0}{1+z^2} = 0; \qquad \dot{\overline{{v}^{(H)}_2}} - \frac{v^{(H)}_1 v_3^0}{1 + z^2} = 0; \qquad v^{(H)}_1(0) = v_1^0; \qquad v^{(H)}_2(0) = v_2^0,
\]
with the function $z$ defined in \eqref{eq:pars_defz}.

The closed set of dual E-L systems can be written down for \eqref{eq:pars_ex_2} as well as the case when $Q$ is left free, but this is not instructive. The main point of the above two examples is to show through explicit computation of a non-trivial example that, in a formulation in which an additive constraint force is added to the equations of unconstrained motion to account for the presence of some kinematic constraints subject only to the power of constraint forces being non-negative, uniqueness of solutions, even for short times, is not to be expected.

\section{A resolution}\label{sec:resol}
The apparent non-uniqueness in the nature of the motion (as well as the constraint force), as evidenced from the examples considered in this paper, is certainly disconcerting (but not entirely unanticipated, cf. Sec.~\ref{sec:intro_constr}). Formulations with equal number of variables and equations have been demonstrated, but uniqueness of solutions can still be an issue, even with restrictions imposed on working of constraint forces that ensure they are dissipative or conservative. Within this general setting, some formulations may deliver uniqueness but then the question of which formulation is `best,' and why, needs to be explored. In this context, the variational principles we suggest allows the use of different choices of $H$ - both through its functional form and its dependence on the `base states,' the collection of specified functions of time specified by overhead bars - as a selection criterion for solutions to the primal problem of constrained particle dynamics.

In closing, we consider the following conceptually minimal generalities free of ad-hoc physical choices, but not necessarily implementable in practical terms, especially for analytical work in terms of explicit formulae. 
For the constrained system of $2d . N$ degrees of freedom (including both position and velocities) with $M$ constraints written in the form
\begin{equation}\label{eq:constr_eliminate}
\begin{aligned}
    \dot{x}_i & = v_i, \qquad i = 1, \ldots, d.N \\
    \dot{v}_i & = F_i(x,v,t) + F_i^{(c)}, \qquad i = 1, \ldots, d.N \\
      0 & = g^\alpha(x,v,t), \qquad \alpha = 1,\ldots, M\\
     v_i^0 & = v_i(0); x_i^0  = x_i(0) \qquad \mbox{specified satisfying } 0 = g^\alpha\left(x^0, v^0,0 \right),
\end{aligned}
 \end{equation}
 assume, as an application of the implicit function theorem, that, in a small neighborhood of the initial condition at least, the array $v$ can be split into two parts and written, by renumbering if necessary, as
 \[
 v = \left(v^{(r)}, v^{(s)} \right)
 \]
such that there exists a function
\[
\hat{v}^{(s)}\left(x, v^{(r)},t \right)
\]
satisfying
\[
g^\alpha \left(x, v^{(r)}, \hat{v}^{(s)}\left(x, v^{(r)},t \right), t \right) = 0.
\]
The number of degrees of freedom in the splitting of $v$ depends on the maximal rank of the matrix $\frac{\p g^\alpha}{\p v_i}(x^0, v^0, 0)$; let this maximal rank be $K \leq M$. We include holonomic constraints in the treatment by formally differentiating such constraints w.r.t time and considering them in `rate' form. The array $v^{(s)}$ may be indexed as $v^{(s)}_{i_s}, i_s = 1, \ldots, K$.  Assuming we have reordered the list $( (x_i, v_i), i = 1,\ldots,d.M)$ as 
\[
\left(x^{(s)}, x^{(r)}, v^{(s)}, v^{(r)} \right) :=  \left( \left(x_{i_s}, x_{i_r}, v_{i_s}, v_{i_r} \right)| \quad i_s = 1,\ldots K, \quad i_r = K+1, \ldots, d.M \right),
\]
and using the notation
\[
x = \left( x^{(s)}, x^{(r)} \right), \quad F = \left( F^{(s)}, F^{(r)} \right),\quad F^{(c)} = \left( F^{(c)(s)}, F^{(c)(r)} \right),
\]
one solves the reduced system
\begin{equation}\label{eq:constr_red_dyanmics}
    \begin{aligned}
        \dot{x}^{(r)}(t)  & = v^{(r)} (t) \\
        \dot{x}^{(s)}(t)  & = \hat{v}^{(s)}\left(x(t),v^{(r)}(t),t \right) \\
        \dot{v}^{(r)}(t)  & = F^{(r)}\left(x(t), \hat{v}^{(s)}\left(x(t), v^{(r)}(t), t \right),v^{(r)}(t),  t\right)\\
        \left(x, v^{(s)}, v^{(r)} \right)(0)  & = \left(x^0, v^{(s)0}, v^{(r)0} \right) \quad \mbox{specified satisfying } \\
        g^\alpha\left(x^0, v^{(s)0}, v^{(r)0}, 0\right) & = 0, \qquad \alpha = 1,\ldots, M \\
    \end{aligned}
\end{equation}
and obtains the solution to the full system by the following evaluation of the constraint forces:
\begin{equation*}
\begin{aligned}
     F^{(c) (r)}(t) & = 0 \\
        F^{(c)(s)}(t) & := \left.  \frac{d}{dt}\left(\hat{v}^{(s)}\left(x(\cdot),v^{(r)}(\cdot), \cdot \right) \right)  \right|_t - F^{(s)}\left(x(t), \hat{v}^{(s)}\left(x(t), v^{(r)}(t), t \right),v^{(r)}(t),  t\right)
\end{aligned}
\end{equation*}
(in the neighborhood where $\hat{v}^{(s)}$ is defined). In order to compute orbits involving states not contained in the domain, say $\Omega_0$, of the function $\hat{v}^s$ corresponding to the specified initial condition in \eqref{eq:constr_eliminate}, the `initial condition' can be reset to the attained state when orbits reach states at or near the boundary of $\Omega_0$, and the scheme above repeated.

The above considerations show the minimal constraint reactions that can be in play in a constrained particle system. No extraneous, ad-hoc, assumptions have been made here about the nature of the constraint forces and, under $C^1$ smoothness of the functions $F$ and $\hat{v}^{(s)}$, local in time uniqueness of solutions is also expected.

We make two observations:
\begin{itemize}
    \item Generating the function $\hat{v}^{(s)}$ is a non-trivial matter, but may not be impenetrable in a computational approximation scheme, although definitely computationally expensive. Within a computational setting, the local nature of the definition of $\hat{v}^{(s)}$ is also not a fundamental barrier to computation of approximate global trajectories of the constrained system.
    \item The methodology for generating dual action principles we have proposed applies seamlessly to the system \eqref{eq:constr_red_dyanmics}.
\end{itemize}
\section{Concluding remarks}\label{sec:concl}

We have developed a method that takes as input a broad class of equations of motion, including those with kinematical constraints that are nonlinear in the component velocities as well as those involving non-conservative forces, and transforms them into a dual system that is governed by a Lagrangian depending on a functional parameter. The dual system can, in many cases, be reformulated in a Hamiltonian framework, giving rise to constants of motion. Our method is applicable to PDE systems and, we expect, should make it possible to extend the method of  path integrals to a wider class of physical systems. More speculatively, we hope to connect with  astrophysical applications as in \cite{goldberger2006dissipative}.

\section*{Acknowledgments}
This work was supported by the Simons Pivot Fellowship grant \# 983171. We thank G.~J.~Wang for bringing the paper \cite{dettmann1996hamiltonian} to our attention.
\bibliographystyle{alpha}\bibliography{particle}

\end{document}